\documentclass[a4paper,11pt]{article}
\usepackage[english]{babel}
\usepackage[utf8]{inputenc}
\usepackage{fancyhdr}
\usepackage{xcolor,colortbl}
\usepackage{framed}
\usepackage{hyperref}
\usepackage[compact]{titlesec}

\usepackage{pdfpages}
\usepackage{authblk}

\usepackage{graphicx}

\usepackage{siunitx}
\usepackage{subcaption, mwe}
\usepackage{mhchem}

\begin{document}

\title{Stable High Power deep-UV Enhancement Cavity in Ultra High Vacuum with Fluoride Coatings}

\author[1]{Zakary Burkley}
\author[1]{Lucas de Sousa Borges}
\author[1]{Ben Ohayon}
\author[2]{Artem Golovozin}
\author[1]{Jesse Zhang}
\author[1]{Paolo Crivelli}
\affil[1]{Institute for Particle Physics and Astrophysics, ETH Z\"urich, CH-8093 Z\"urich, Switzerland}
\affil[2]{Lebedev Physical Institute, Moscow, Russia}

\maketitle


\begin{abstract}
We demonstrate the superior performance of fluoride coated versus oxide coated mirrors in long term vacuum operation of a high power deep-ultraviolet enhancement cavity. In high vacuum (\SI{e-8}{\milli\bar}), the fluoride optics can maintain up to a record-high \SI{10}{\watt} of stable intracavity power on one hour time scales, whereas for the oxide optics, we observe rapid degradation at lower intracavity powers with a rate that increases with power. After observing degradation in high vacuum, we can recover the fluoride and oxide optics with oxygen; however, this recovery process becomes ineffective after several applications. For fluoride coatings, we see that initial UV conditioning in an oxygen environment helps to improve the performances of the optics.  In oxygen-rich environments from $\sim$\SIrange{e-4}{1}{\milli\bar}, the fluoride optics can stably maintain up to \SI{20}{\watt} of intracavity power on several-hour time scales whereas for the oxide optics there is immediate degradation with a rate that increases with decreasing oxygen pressure.
\end{abstract}

\section{Introduction}

Continual innovation of deep-ultraviolet (deep-UV) laser technology is beneficial for precision scientific measurement, materials processing, biomedicine, and many other industrial and medicinal applications \cite{savage2007ultraviolet, lu2020deep, smith2020low}. 
Understanding the performance of optics at high average deep-UV powers is of growing importance. Both picosecond pulsed and continuous wave laser systems below \SI{250}{\nano\meter} operating at Watt-level average powers have been demonstrated recently \cite{willenberg2020high, turcicova2019, Burkley19}. Numerous applications would immediately gain from increased stability at these powers such as laser cooling of \ce{AlF} at \SI{227.5}{\nano\meter} \cite{truppe2019spectroscopic} and mercury at \SI{253.7}{\nano\meter} \cite{zhao2017high}, as well as two-photon cooling of hydrogen at \SI{243}{\nano\meter} \cite{cooper2018}.  A laser-based parity measurement in hydrogen \cite{rasor2020laser} would also benefit from increased deep-UV Rabi frequencies, as would Rydberg excitation for fast entanglement of strontium ions \cite{zhang2020submicrosecond} or studies of strong long-range interactions in a lithium optical lattice \cite{guardado2020quench} at \SI{243}{\nano\meter} or \SI{231}{\nano\meter}, respectively. Anti-matter wave interferometry with positronium could also be envisaged \cite{Oberthaler:2002qd}.

One particular area of research that profits notably from advances in deep-UV radiation sources is precision two-photon spectroscopy of simple and exotic atoms. For example, comparison of the 1S-2S transition in hydrogen \cite{parthey2011improved} and anti-hydrogen \cite{ahmadi2018} has tested CPT invariance at a level of \num{2e-12}, and requires \SI{243}{\nano\meter} radiation. At \SI{205}{\nano\meter}, differing measurements on the 1S-3S transition of hydrogen \cite{fleurbaey2018, grinin2020} has contributed notably to the proton radius puzzle. A new experiment underway on the 1S-2S transition in Muonium at \SI{244}{\nano\meter} promises a substantial improvement in the determination of the muon mass and tests of lepton universality \cite{crivelli2018}; at \SI{252.5}{\nano\meter}, precision two-photon spectroscopy in Xenon \cite{altiere2018} has recently been demonstrated for future neutron EDM measurements at TRIUMF \cite{picker2017minuscule}. 

While the high precision advantage of two-photon spectroscopy comes from its removal of the first-order Doppler effect and access to long-lived states, this second order process suffers from low cross-sections. This can be overcome with the high peak power offered through pulsed laser systems, but introduces many systematics---these can be removed through use of a continuous wave (CW) laser enhanced in an optical cavity. While such enhancement cavities have been built, e.g., \cite{parthey2011improved, ahmadi2018, altiere2018}, they operate at modest intracavity powers (\SI{<1}{\watt} CW), and are limited to low buildup by the available reflectivity of deep-UV optics. Impressively, up to \SI{30}{\watt} of CW intracavity power has been demonstrated, but required \SI{0.67}{\milli\bar} of oxygen on the mirrors to prevent UV-assisted degradation \cite{cooper2018}. This degradation is attributed to surface oxygen depletion and/or hydrocarbon contamination \cite{kunz2000experimentation, hollenshead2006, gangloff2015}. In principle, both of these effects are reversible by keeping \ce{O2} present on the mirrors, as was required in \cite{cooper2018, grinin2020, altiere2018}. However, such a solution is not ideal. In \cite{altiere2018}, the precision of the measurement was limited by pressure shifts introduced by the required \ce{O2} level. In \cite{cooper2018}, reaching a vacuum of \SI{5e-8}{\milli\bar} for spectroscopy required a technically demanding differential pumping setup with geometric constraints that significantly limited the intracavity power \cite{cooper2020}. Furthermore, implementing such pumping requirements would be challenging in the setups required to study anti-hydrogen \cite{ahmadi2018} and muonium \cite{crivelli2018} where complicated beamline and cryogenic environment constraints  exist.
For these reasons, a stable deep-UV cavity operating with Watt-level circulating powers in high vacuum or reduced oxygen pressures would be of great interest to the laser spectroscopy community.  

The previous enhancement cavities that observed degradation utilized oxide coatings for their high reflective mirrors. To better understand the degradation process, an interesting alternative are fluoride coatings. With no oxides present, the latter should be insensitive to surface oxygen depletion. Furthermore, low fluence pulsed studies of fluoride coatings show an extremely high durability with initial UV conditioning, an effect not observed with oxide coatings \cite{eva1996laser,heber1999changes}. Therefore, in this work, we test the performance of both fluoride-based and oxide-based  coatings at intra-cavity powers from \SIrange{1}{10}{\watt} at \SI{244}{\nano\meter} in high vacuum, and \SIrange{15}{20}{\watt} in various oxygen pressures. At high powers, we observe superior performance in the stability of the fluoride coated optics relative to the oxide coated optics.

\begin{figure}[h]
  \centering
  \includegraphics[width=\textwidth]{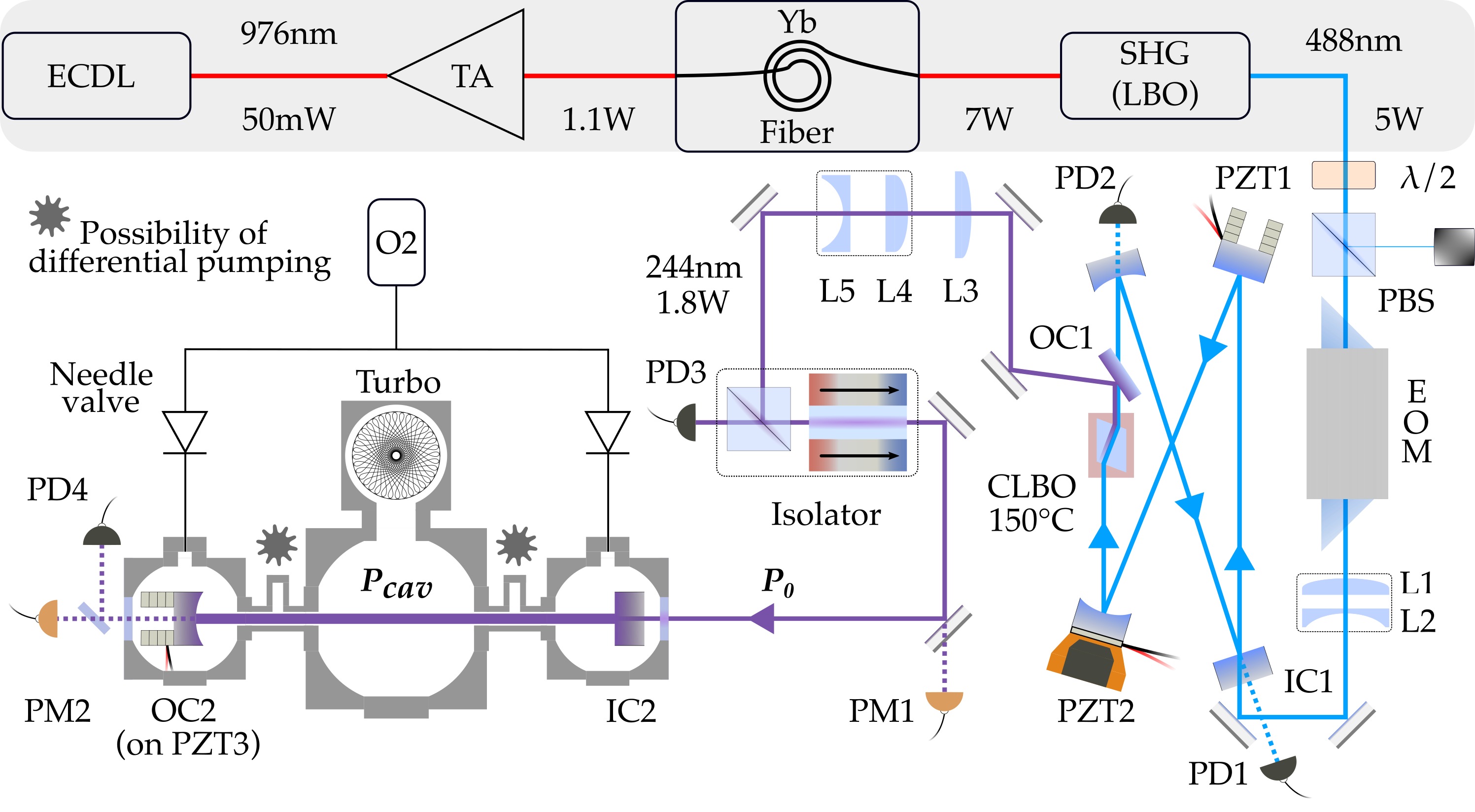}  
  \caption{Experimental setup: \SI{244}{\nano\meter} radiation is generated through a frequency quadrupled ytterbium-doped fiber amplified laser system. This radiation is then coupled into a vacuum enhancement cavity. Extended cavity diode laser (ECDL), tapered amplifier (TA), second harmonic generation (SHG), lithium triborate (LBO), polarizing beam splitter (PBS), electro-optic modulator (EOM), L1-L5 (mode matching lenses), photodiode (PD), power monitor (PM), caesium lithium borate (CLBO), input coupler (IC), output coupler (OC), piezoelectric transducer stack (PZT).} 
  \label{fig:setup}
\end{figure}

\section{Experimental setup}

Inspired by recent demonstrations of high-power Yb-fiber amplifiers \cite{burkley2017yb, wu18, Burkley19} near \SI{976}{\nano\meter}, the system utilizes a frequency doubled fiber-amplifier developed in collaboration with Toptica for generation of the fundamental IR radiation at \SI{976}{\nano\meter} and second harmonic generation to \SI{488}{\nano\meter}. The system can deliver up to \SI{7}{\watt} of IR power and \SI{5}{\watt} in the blue; with this, up to \SI{1.8}{\watt} of \SI{244}{\nano\meter} power can be generated in a "home-built" resonant doubling stage. 

The second harmonic enhancement cavity that generates 244 nm radiation uses a standard bow tie geometry and is maintained on resonance using the Pound-Drever-Hall (PDH) locking technique \cite{drever1983laser}. A slow piezoelectric (PZT) stack (CTS NAC2123-H20-C01) and fast PZT chip (CTS NAC2122-C04) are used for active length stabilization, with the latter mounted using the technique described in \cite{briles2010simple}. The cavity mirrors (LaserOptik Gmbh) consist of a $2.5 \%$ transmission input coupler and three HR mirrors at \SI{488}{\nano\meter} ($<99.8\%$); two of these mirrors have a radius of curvature of 200 mm to focus the beam within the doubling crystal with a waist of \SI{47}{\micro\meter} by \SI{44}{\micro\meter} . The doubling occurs in a \SI{10}{\milli\meter} long Brewster-cut Caesium Lithium Borate (\ce{CLBO}) crystal (Oxide) in a Type-1 critical phase matching configuration ($\theta = 76.4 ^{\circ}, \phi = 45 ^{\circ}$). The hygroscopic crystal is kept at \SI{150}{\celsius} in an \ce{Al} oven and flushed continuously with oxygen. A dichroic Brewster plate (Spectral Optics) couples the UV radiation out of the cavity. 

\begin{figure}[h]
  \centering
  \includegraphics[width=\textwidth]{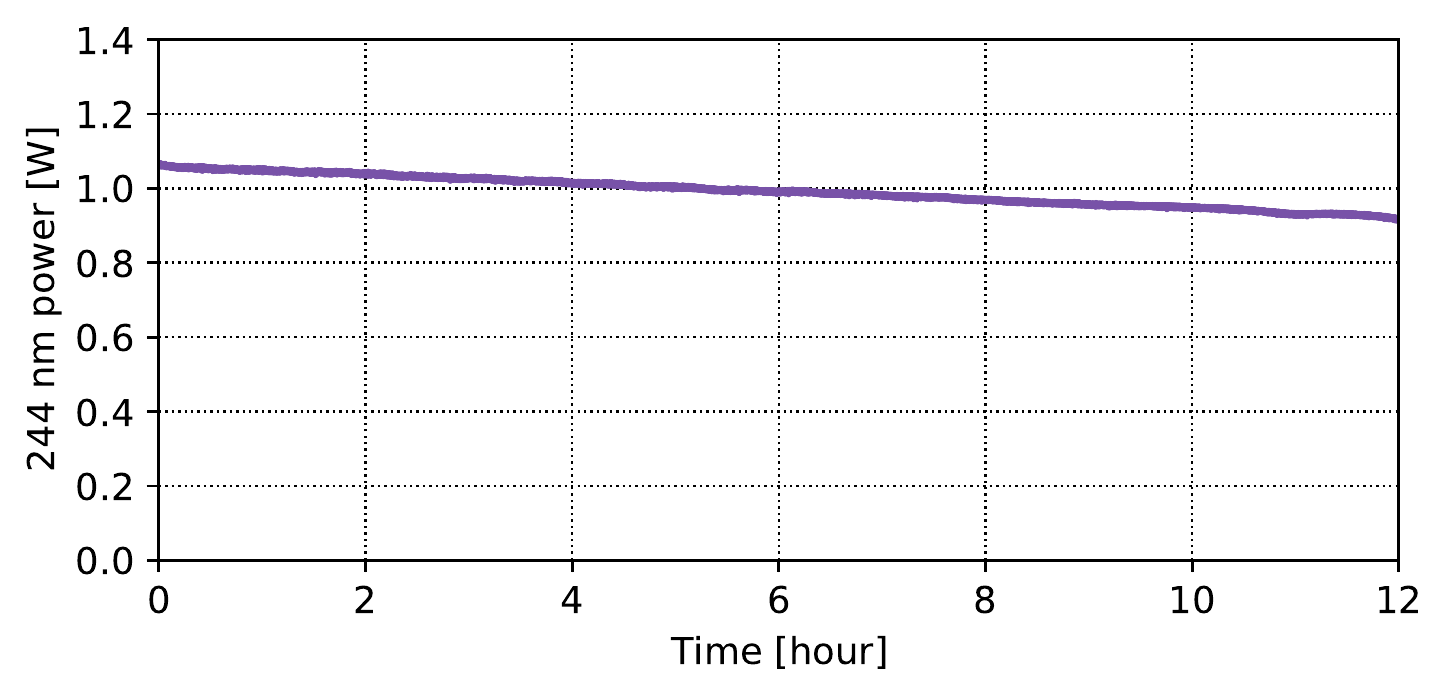}  
  \caption{Performance of the \SI{244}{\nano\meter} laser system over long time scales. The laser can output \SI{1}{\watt} of \SI{244}{\nano\meter} power over several hour time scales without degradation of the \ce{CLBO} crystal. We attribute the slow degradation observed to damage of the output coupler.}
  \label{fig:244out}
\end{figure}

As was shown in \cite{Burkley19}, the \ce{CLBO} can output \SI{>1}{\watt} of radiation on the one hour time scale with no evidence of degradation. Enabled by the increased stability in the IR and blue stages of our laser system, we have observed similar powers on significantly longer time scales (See Fig. \ref{fig:244out}). We measure a slight degradation of $\sim$1-2$\%$ per hour in total power, which we attribute to damage on the \SI{244}{\nano\meter} coating of the output coupler. This damage is visible by eye upon inspection, and the power can be recovered by moving to a new spot on the output coupler.

The \SI{244}{\nano\meter} light is enhanced in a linear, vacuum cavity composed of two calcium fluoride (\ce{CaF2}) substrates coated with fluoride through electron beam evaporation (LaserOptik Gmbh). The input coupler (IC) is flat with a measured transmission of $0.024$, and the curved (radius of curvature = \SI{4}{\meter}) high reflector (HR) has a manufacturer specified reflectivity $> 98.5\%$. The mirrors are \SI{0.50}{\meter} apart, and the Gaussian beam inside the cavity is approximately collimated with a beam waist of \SI{\sim 0.3}{\milli\meter}. One spherical lens and two cylindrical lenses (to compensate for walk-off in the \ce{CLBO}) are used for mode matching into cavity. When optimally aligned, measurements of the rejection signal ratio indicate input coupling efficiency of $80\%$. We isolate the back-reflected power using a polarizing beam-splitter and a Faraday rotation stage. A portion of this back-reflected signal is used for PDH locking, where the HR is mounted on an annular PZT stack (CTS NAC2123-H20-C01) in the method described in \cite{chadi2013} for active stabilization. 

The transmitted power through the HR is used to monitor the intracavity power with a calibrated UV thermal power meter (Thorlabs S401C). This, combined with the measured HR transmission ($T_{\text{HR}} = 0.0012$), yields the intracavity power, $P_{cav}$. As seen in Fig. \ref{fig:setup}, the input power, $P_{o}$ is monitored with a sensitive, calibrated UV photodiode (Thorlabs S120VC) placed directly after the last mirror before the vacuum enhancement cavity. The mirror is backside polished, and the leakage light is calibrated to the UV power entering the cavity. In this way, both the input power and intracavity power are monitored non-destructively, enabling in situ measurement of the cavity power enhancement $\beta = P_\text{\it{cav}}/P_\text{\it{o}}$ in a way which is independent of any power drifts from the UV laser system leading to the cavity. 

The cavity is operated in a variable oxygen purge mode or ultra high vacuum mode. In the \ce{O2} mode, two leak valves are adjusted to maintain a fixed flow of fresh \ce{O2} (6.0 grade) on the cavity mirrors. For high \ce{O2} pressures of \SI{e-1}{\milli\bar} or more, only a roughing pump is used. For \ce{O2} pressures at or below \SI{e-3}{\milli\bar}, a turbo pump is used to maintain the pressure equilibrium. The lowest equilibrium pressure of fresh \ce{O2} we can confidently maintain is \SI{e-4}{\milli\bar}, limited by the precision of the leak valves. In the UHV mode, there is no oxygen purge, and the pressures reach below \SI{e-8}{\milli\bar}. To limit possible hydrocarbon contaminants, we minimize use of any plastics and only use UHV compatible components. The vacuum windows are sealed with indium instead of O-rings, the PZT is UHV compatible, the HR is mounted to the PZT with Torr seal, and the PZT is mounted following the design in \cite{chadi2013} using PEEK screws, as PEEK is known to have low outgassing rates. All other elements in the system are either stainless steel, copper, or aluminum.  

\section{Results and discussion}

We measured the power enhancement factor $\beta$ versus time for both a \ce{SiO2} and \ce{CaF2} input coupler in high vacuum and different oxygen pressures, with various intracavity powers. The high reflector is \ce{CaF2} for all measurements. 

\begin{figure}[h]
\captionsetup[subfigure]{justification=centering}
\begin{subfigure}{0.5\textwidth}
  \centering
  \includegraphics[width=\textwidth]{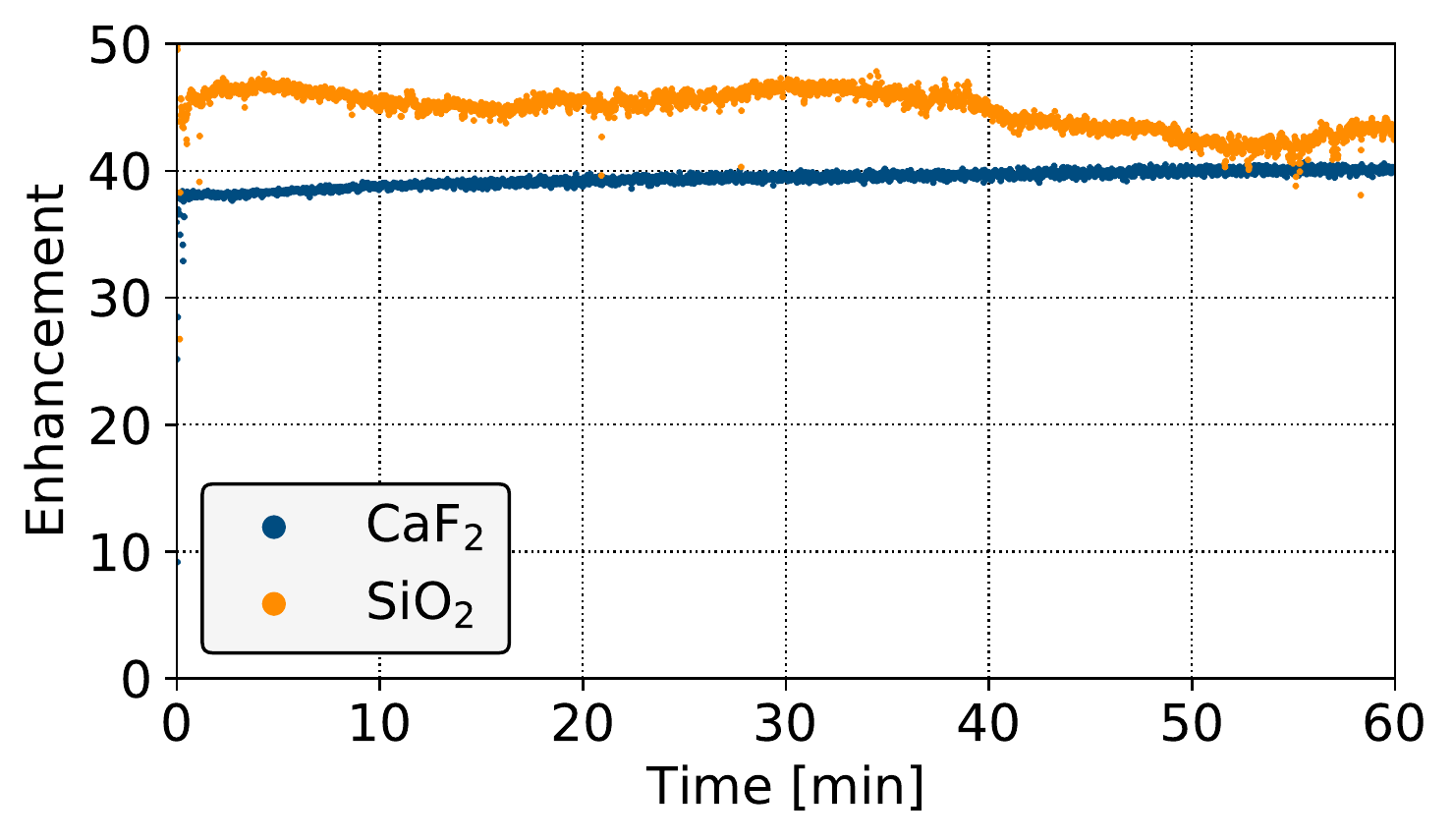}  
  \subcaption{ \SI{1.5}{\watt} intra-cavity power for both input couplers.}
  \label{fig:UHVa}
\end{subfigure}
\begin{subfigure}{0.5\textwidth}
  \centering
  \includegraphics[width=\textwidth]{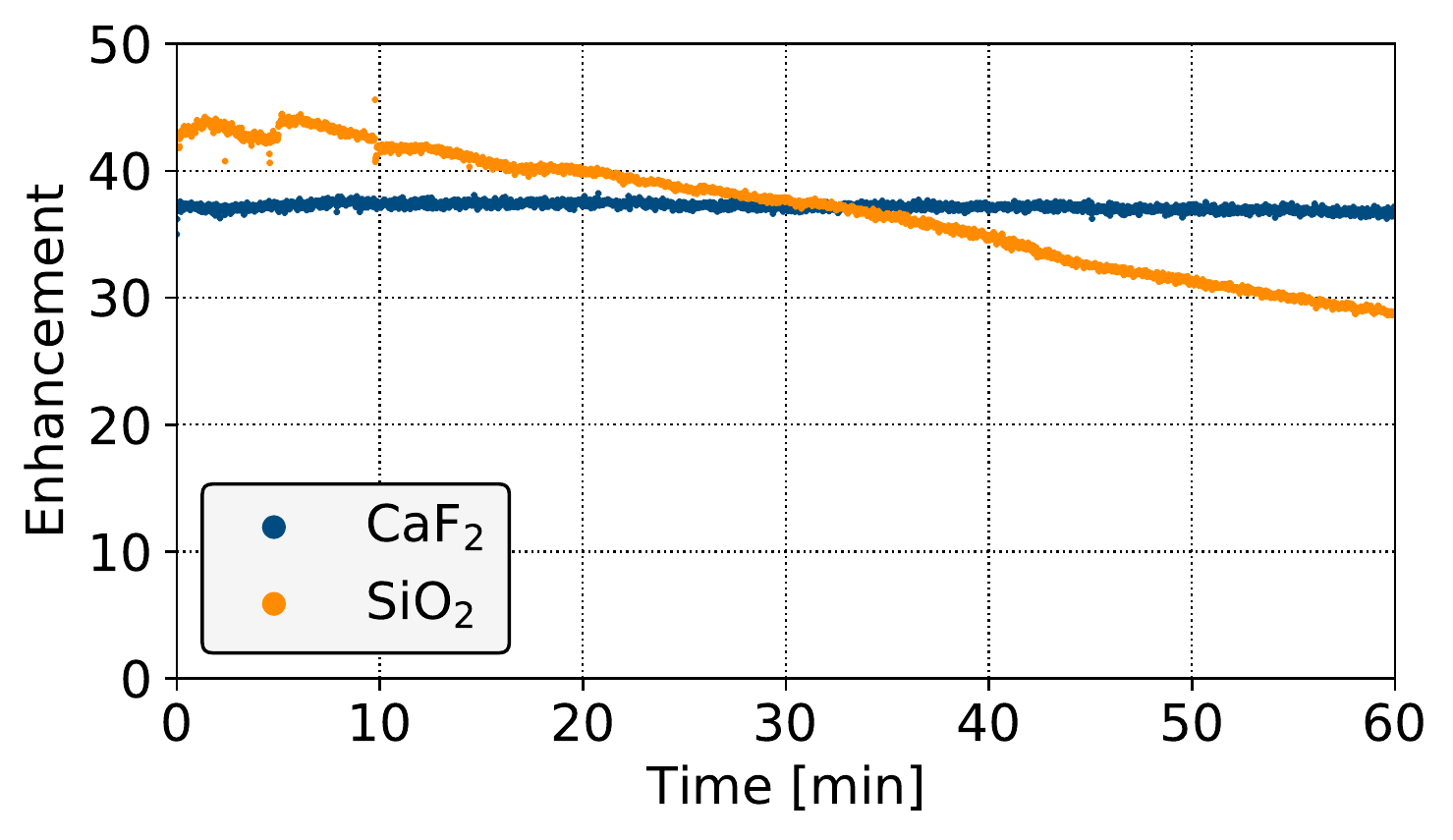}  
  \subcaption{ \SI{3}{\watt} intra-cavity power for both input couplers.}
  \label{fig:UHVb}
\end{subfigure}
\begin{subfigure}{0.5\textwidth}
  \centering
  \includegraphics[width=\textwidth]{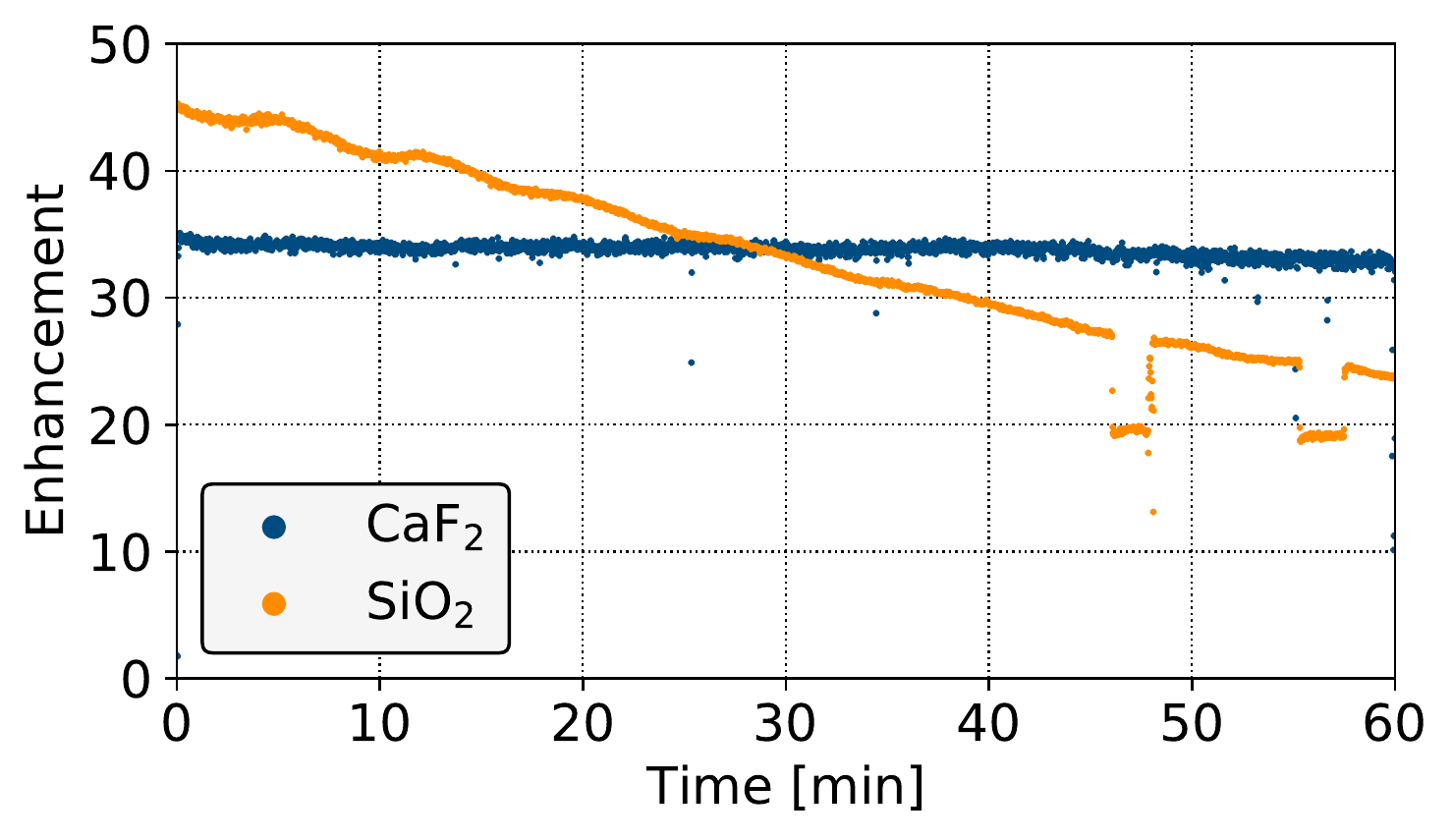}  
  \subcaption{\SI{5}{\watt} intra-cavity power for \ce{CaF2} IC,\\ \SI{5.5}{\watt} start power for \ce{SiO2} IC.}
  \label{fig:UHVc}
\end{subfigure}
\begin{subfigure}{0.5\textwidth}
  \centering
  \includegraphics[width=\textwidth]{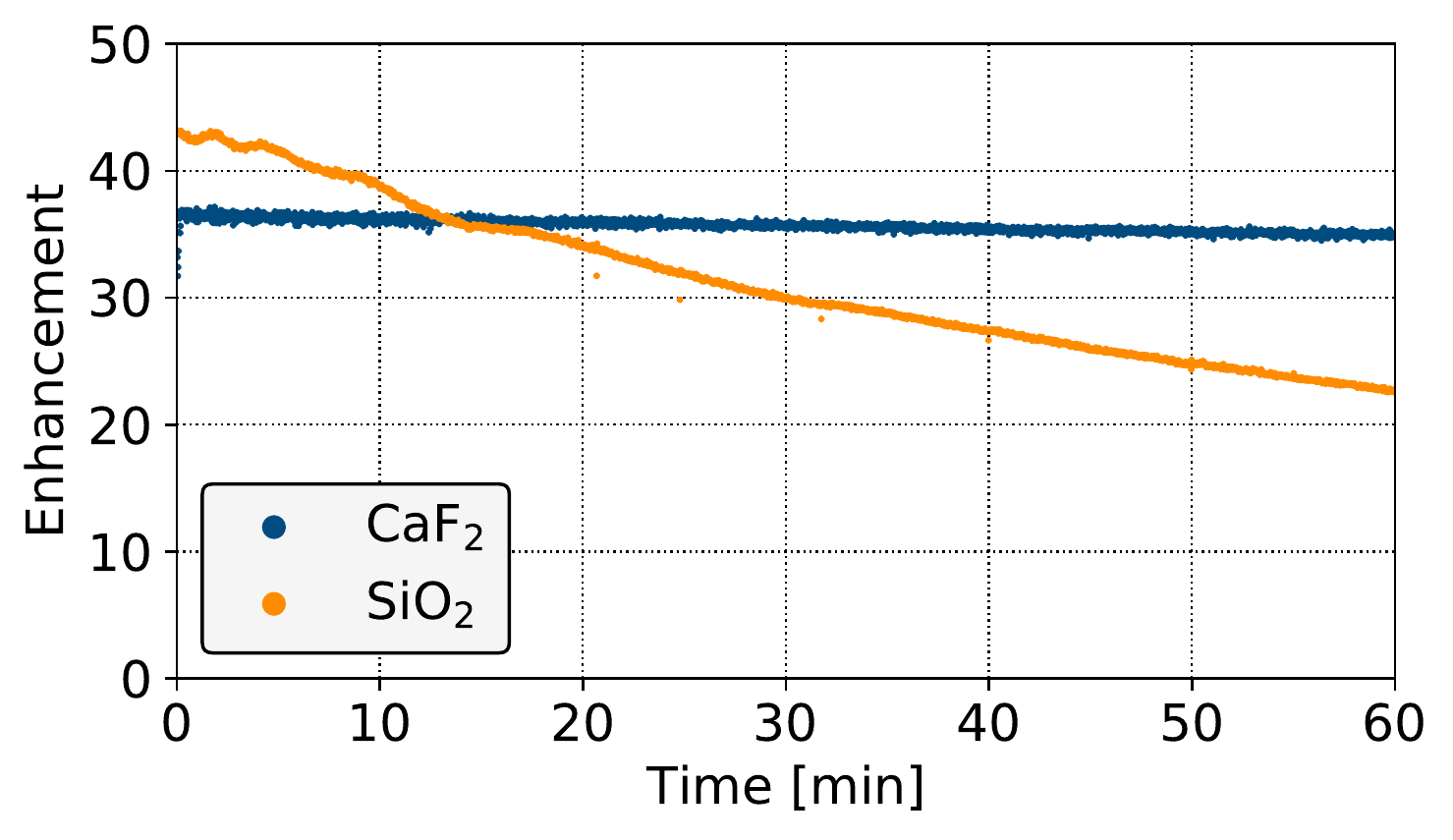} 
  \subcaption{\SI{10}{\watt} intra-cavity power for \ce{CaF2} IC,\\ \SI{9}{\watt} start power for \ce{SiO2} IC.}
  \label{fig:UHVd}
\end{subfigure}
\caption{Enhancement versus time of \ce{CaF2} IC versus \ce{SiO2} IC for different intracavity powers at \SI{e-8}{\milli\bar}.}
\label{fig:UHVdata}
\end{figure}

\begin{figure}
  \centering
  \includegraphics[width=0.75\textwidth]{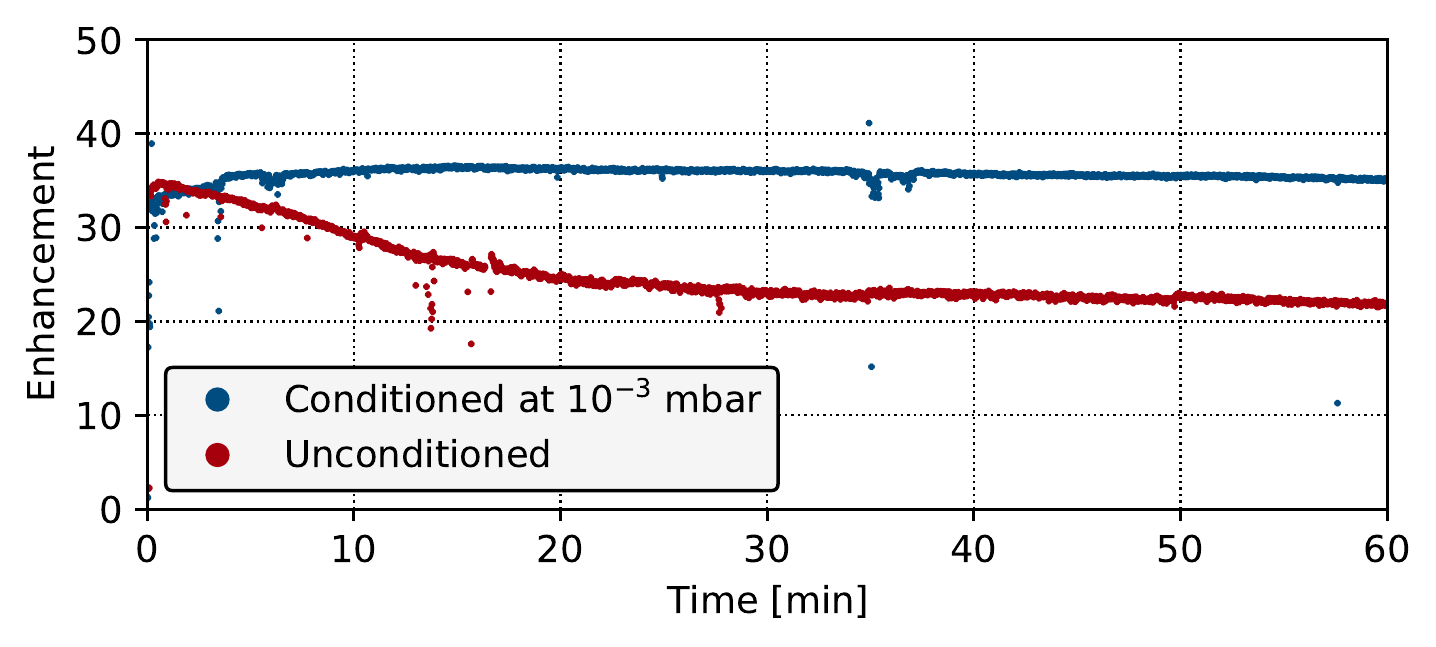}  
  \caption{Enhancement versus time for unconditioned and conditioned \ce{CaF2} IC. Intracavity power of \SI{8}{\watt} for conditioned IC, \SI{10}{\watt} start power for unconditioned IC.}
  \label{fig:conditioning}
\end{figure}

In Figure \ref{fig:UHVdata}, we give the enhancement factor in high vacuum as a function of time for both a \ce{SiO2} and \ce{CaF2} input couplers for different intracavity powers. At low power of \SI{\sim 1.5}{\watt} (Fig. \ref{fig:UHVa}), we observe little or no degradation for both coatings. However, there appears to be slow oscillations of small amplitude, with the \ce{SiO2} input coupler.
As we increase the intracavity powers by inserting more input power (Figs. \ref{fig:UHVb}-\ref{fig:UHVd}), we observe an increased rate of degradation with the \ce{SiO2} IC, limiting the asymptotic power to a few watts. Small amplitude, slow oscillations are also observed in these measurements, and appear to relax on a shorter time scale for increased starting powers. At higher powers, the enhancement factor decreases by approximately a factor of two in one hour, with a rate that decreases as the overall enhancement factor decreases. Similar to what was observed in \cite{cooper2018}, degradation is easily visible on the coating after inspection.
For \ce{CaF2}, the data given in Figs. \ref{fig:UHVa}-\ref{fig:UHVd} was measured continuously while increasing the input power every hour. Furthermore, prior to the high vacuum measurement, the \ce{CaF2} mirror was conditioned in \SI{1e-3}{\milli\bar} of \ce{O2} with \SI{16}{\watt} of intracavity power for several hours. As seen in Figs. \ref{fig:UHVa}-\ref{fig:UHVd}, performance with the \ce{CaF2} IC is stable on one hour timescales with up to \SI{10}{\watt} of intracavity power. A slight decrease in the enhancement from 40 to 34 is visible.  We attribute the drop in enhancement to a slow degradation of the \ce{CaF2} input coupler. Rotating the IC is sufficient to regain enhancement, and upon careful inspection, an inhomogeneity in the coating is visible by eye after running at these high powers in high vacuum for an extended period. Nonetheless, the drop in enhancement of 4$\%$ per hour seen with this IC is an order of magnitude smaller than the   50$\%$ drop seen with the oxide-coated input coupler. As shown in Fig. \ref{fig:conditioning}, we observed that the rate of enhancement decrease for \ce{CaF2} was highly reliant on the initial UV conditioning of the mirrors in an oxygen environment. 

\begin{figure}[h]
  \centering
  \includegraphics[width=0.75\textwidth]{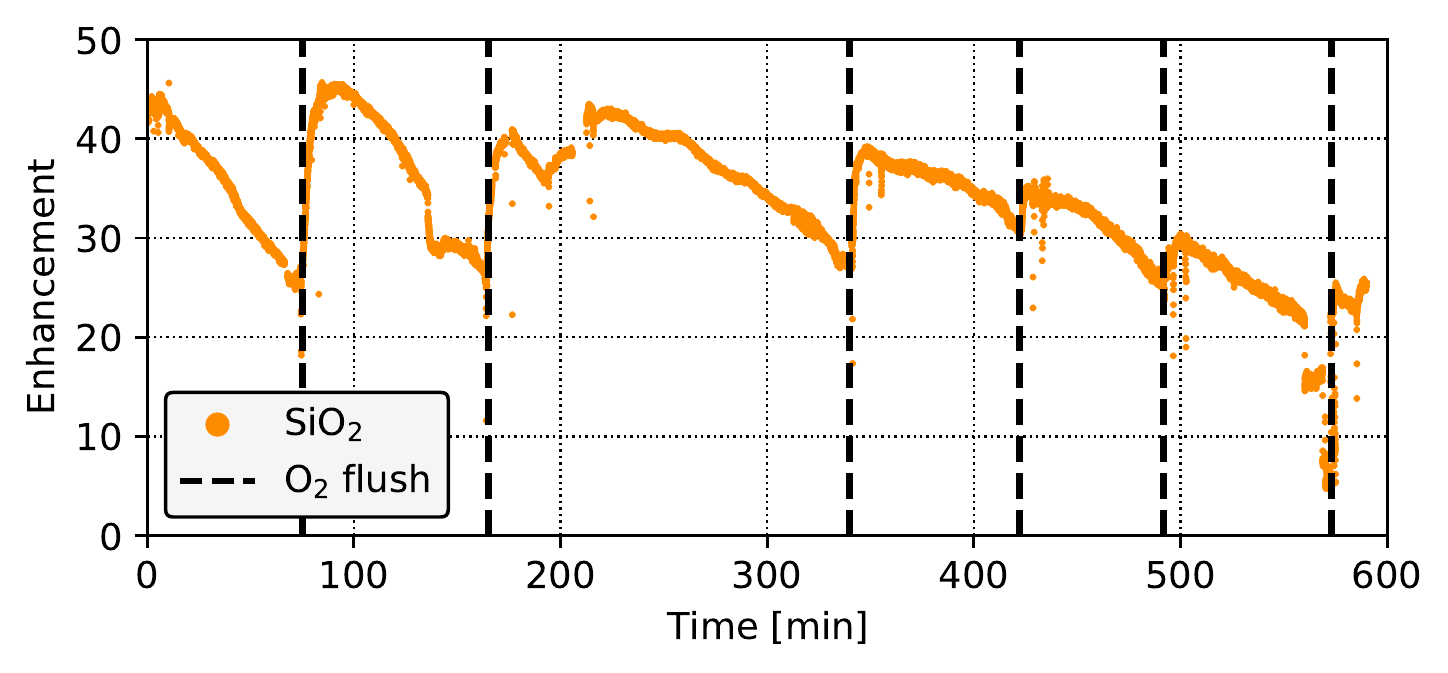}  
  \caption{Oxygen recovery of enhancement at \SI{e-8}{\milli\bar} with \SI{3}{\watt} start power for \ce{SiO2} IC, by momentarily flushing \ce{O2} on the mirrors surface.}
  \label{fig:UHVe}
\end{figure}

We next tested whether higher average intra-cavity powers could be accommodated, relying on the higher start enhancement of the \ce{SiO2} coated IC by introducing intermittent oxygen revivals. As shown in Fig. \ref{fig:UHVe}, in accordance with \cite{gangloff2015, cooper2018}, the \ce{SiO2} optics can be rapidly revived after degradation by flowing oxygen while maintaining a buildup of ultraviolet light in the enhancement cavity. However, even though reviving is possible a few times, eventually the enhancement can no longer be fully recovered. A similar behavior was observed with the \ce{CaF2} input coupler.

\begin{figure}[ht]
\captionsetup[subfigure]{justification=centering}
\begin{subfigure}{0.5\textwidth}
  \centering
  \includegraphics[width=\textwidth]{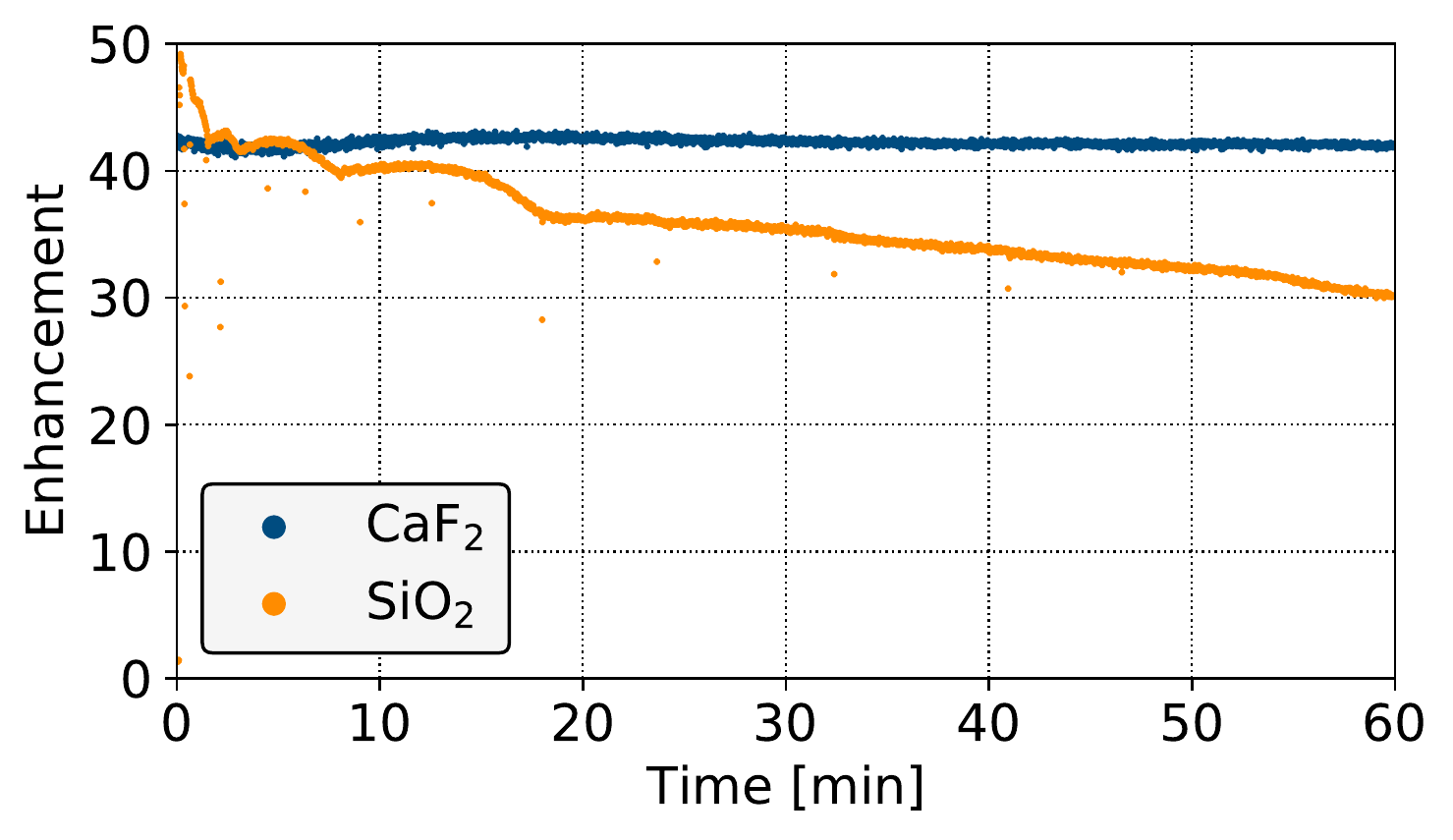}  
  \subcaption{\SI{1}{\milli\bar} of \ce{O2}.\\ \SI{15}{\watt} of intra-cavity power for \ce{CaF2} IC,\\ \SI{17}{\watt} start power for \ce{SiO2} IC.}
  \label{fig:O2a}
\end{subfigure}
\begin{subfigure}{0.5\textwidth}
  \centering
  \includegraphics[width=\textwidth]{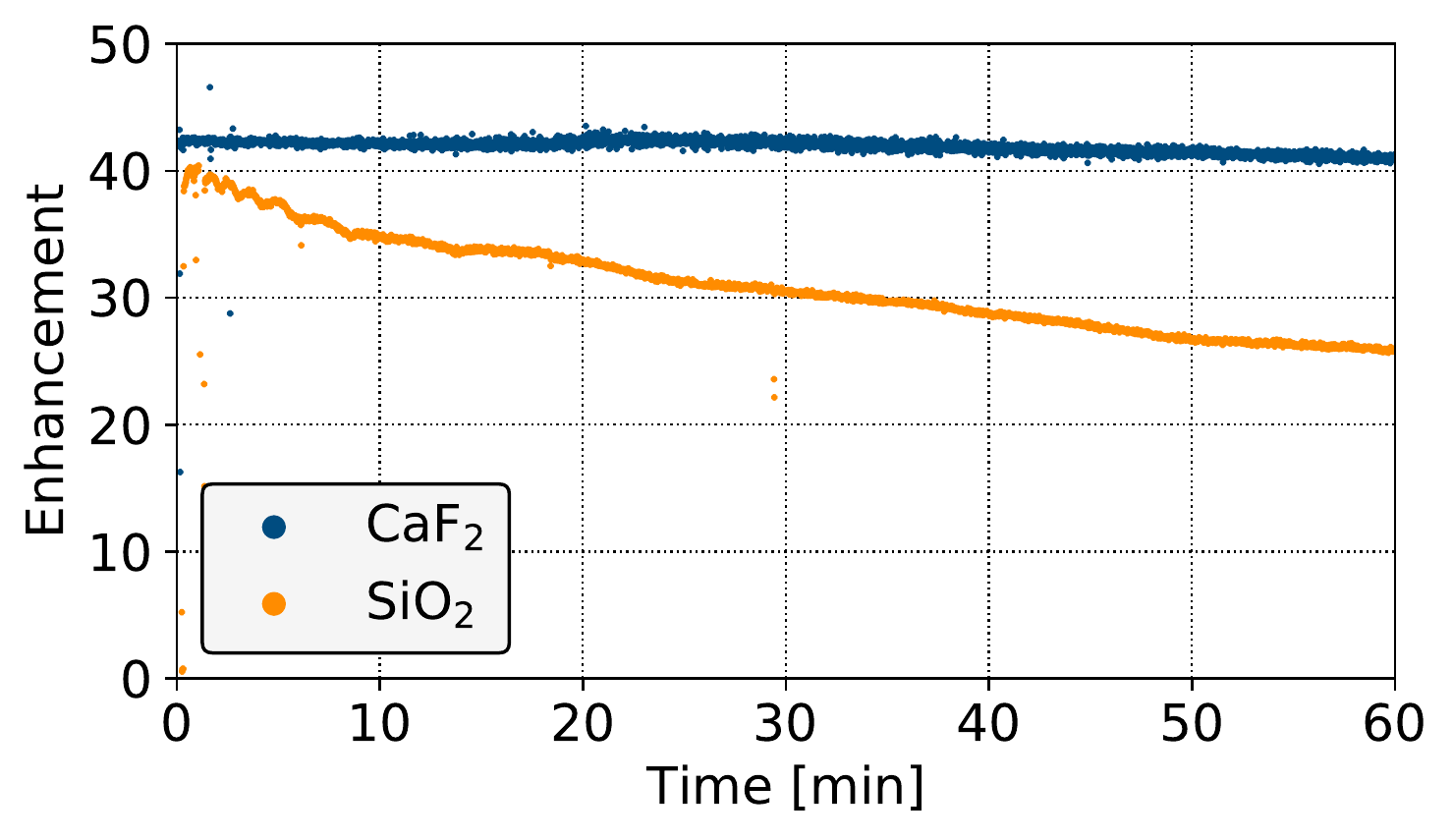}
  \subcaption{\SI{e-1}{\milli\bar} of \ce{O2}.\\ \SI{15}{\watt} intra-cavity power for \ce{CaF2} IC,\\ \SI{12}{\watt} start power \ce{SiO2} IC.}
  \label{fig:O2b}
\end{subfigure}
\begin{subfigure}{0.5\textwidth}
  \centering
  \includegraphics[width=\textwidth]{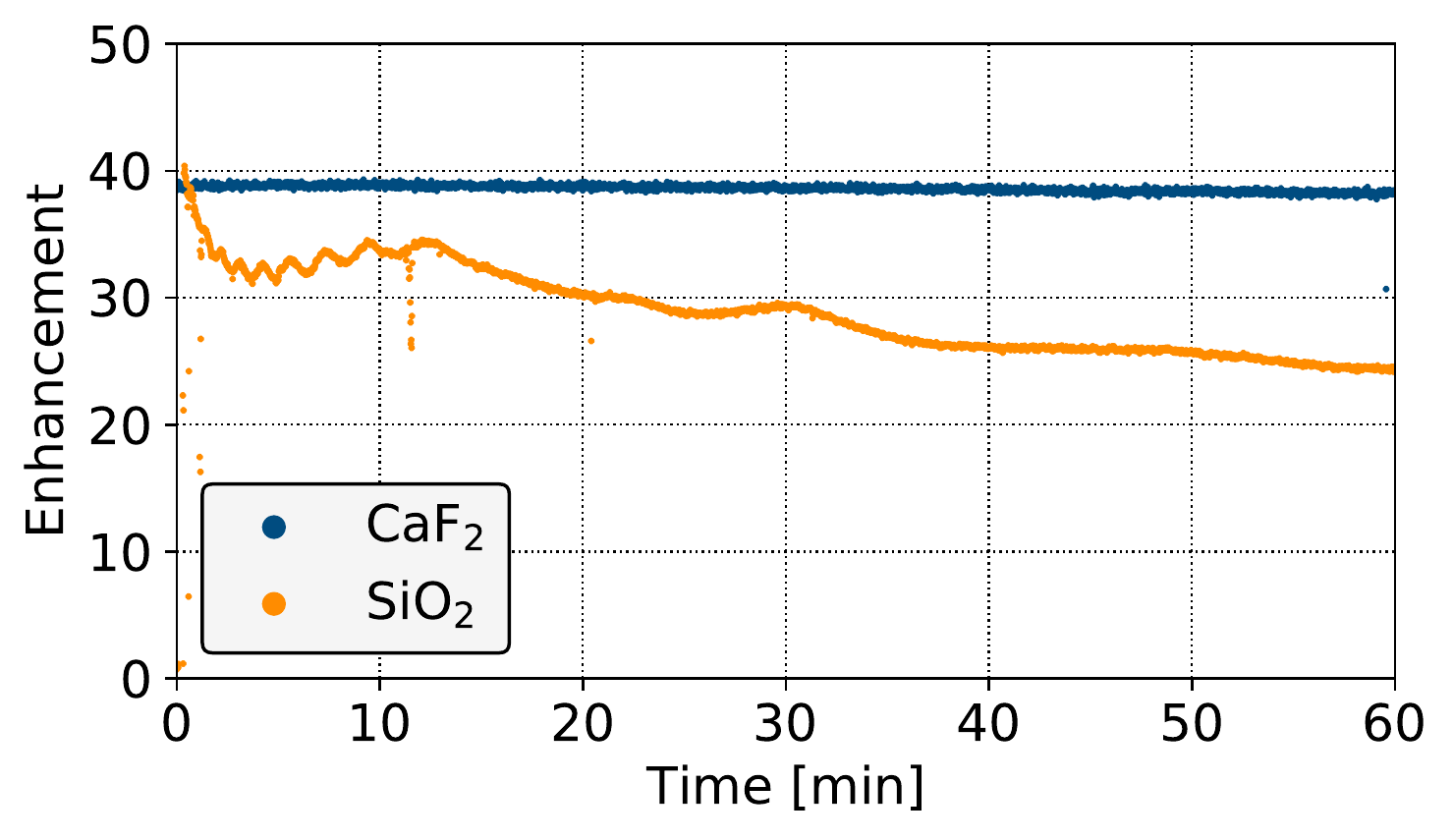}  
  \subcaption{\SI{e-3}{\milli\bar} of \ce{O2}.\\ \SI{18}{\watt} intra-cavity power for \ce{CaF2} IC,\\ \SI{13}{\watt} intra-cavity start power for \ce{SiO2} IC.}
  \label{fig:O2c}
\end{subfigure}
\begin{subfigure}{0.5\textwidth}
  \centering
    \includegraphics[width=\textwidth]{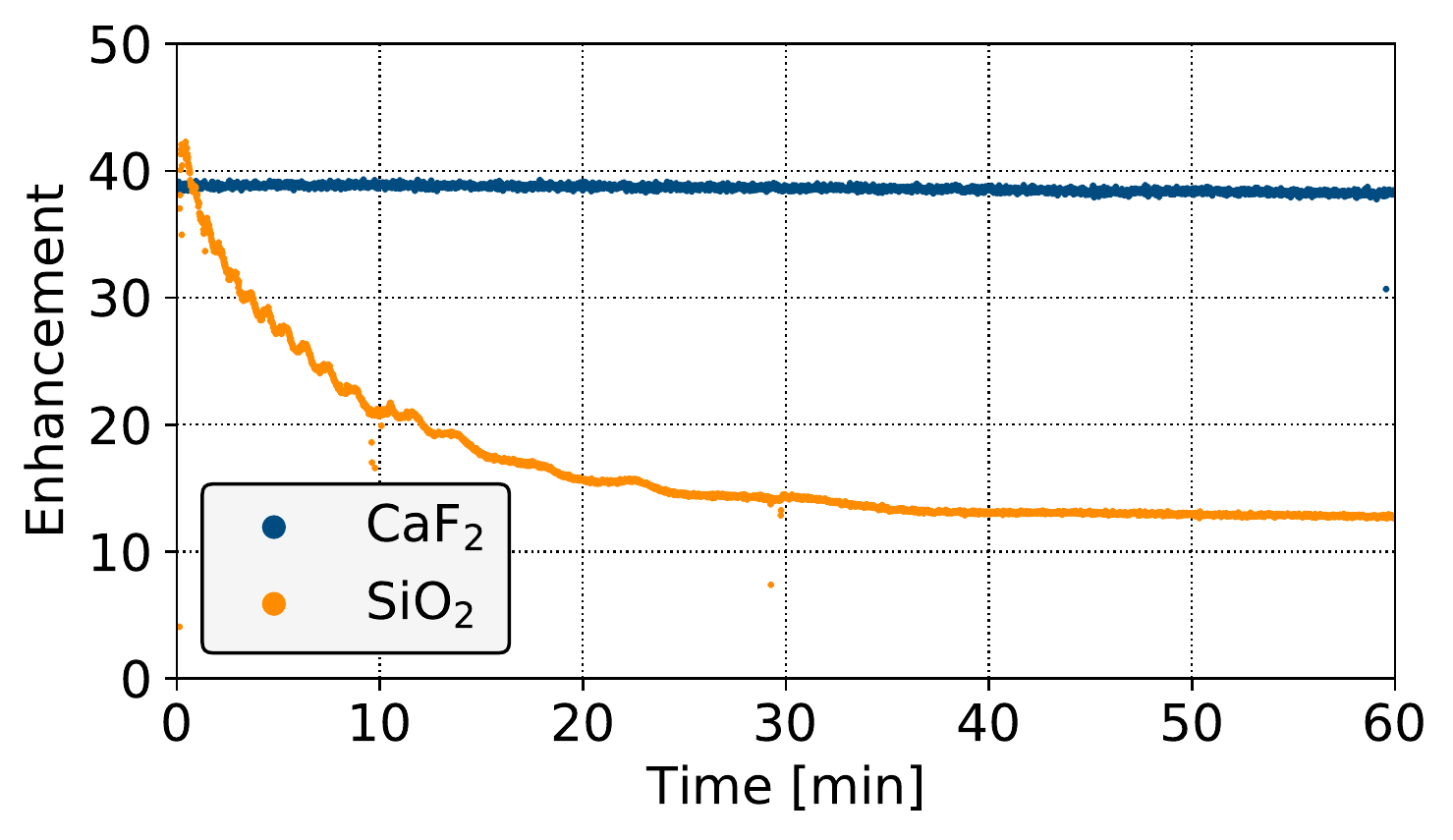}
  \subcaption{\SI{e-4}{\milli\bar} of \ce{O2}.\\ \SI{18}{\watt} intra-cavity power for \ce{CaF2} IC,\\ \SI{12}{\watt} start power for \ce{SiO2} IC.}
  \label{fig:O2d}
\end{subfigure}
\caption{Enhancement versus time of \ce{CaF2} IC versus \ce{SiO2} IC  at different \ce{O2} pressures.}
\label{fig:O2data}
\end{figure}

As observed, an intermittent revival mode with oxygen in high vacuum is not a sustainable option. However, as oxygen revives the input couplers, it is interesting to study their performance in oxygen-rich environments. The performance of the mirrors at high intracavity powers with a continuous oxygen purge of \SIrange{e-4}{1}{\milli\bar} is shown in Fig. \ref{fig:O2data}. With \ce{SiO2}, we observe a rapid initial drop in enhancement, that increases with decreasing oxygen pressure. However, with \ce{CaF2}, we see stable powers over the entire range of oxygen pressures.

\begin{figure}
  \centering
  \includegraphics[width=0.75\textwidth]{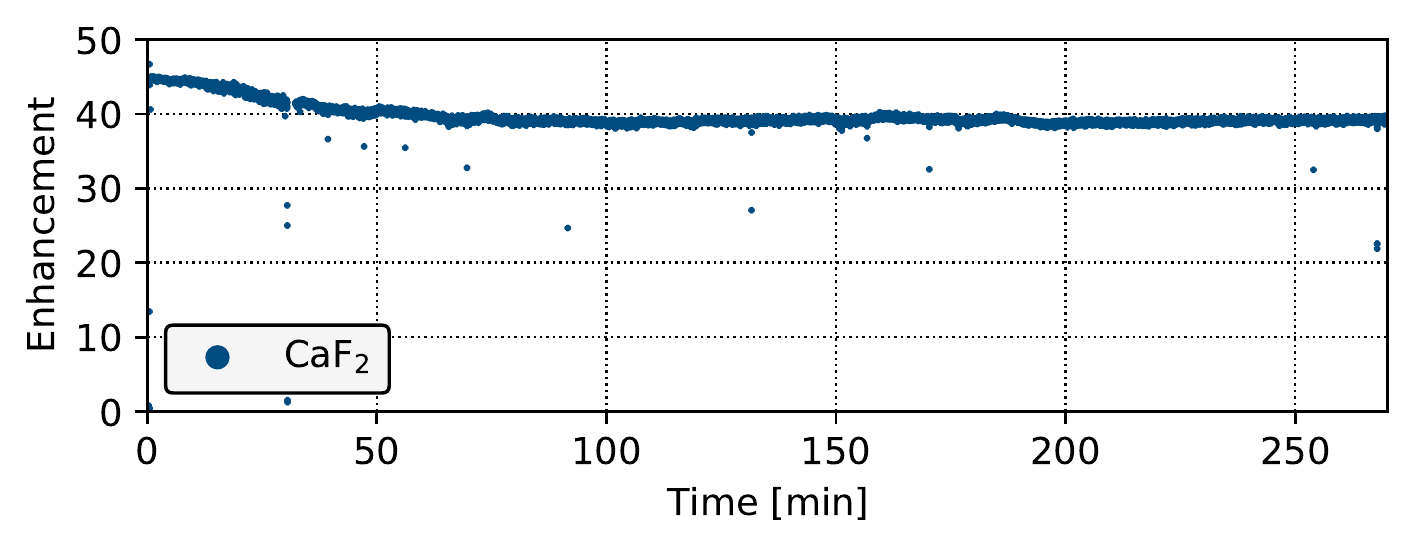}  
  \caption{\SI{e-3}{\milli\bar} of \ce{O2}, \SI{16}{\watt} intra-cavity power for \ce{CaF2} IC.}
  \label{fig:O2e}
\end{figure}

As was the case with Figs. \ref{fig:UHVa}-\ref{fig:UHVd}, Figs. \ref{fig:O2a}-\ref{fig:O2d} were measured continuously for \ce{CaF2}. Unlike the high vacuum case, we do not observe a noticeable drop in enhancement over this four hour time scale, indicating improved performance in oxygen. 
To test this, we moved to a new spot on the \ce{CaF2} IC, and measured for four hours at \SI{e-3}{\milli\bar} as seen in Fig. \ref{fig:O2e}. After an initial slight decrease in enhancement, we observe degradation-free performance over this time scale. We attribute the initial drop in the first 30 minutes to thermal heating of the optics before the enhancement cavity, which can affect the input coupling efficiency. 

To our knowledge, CW intracavity powers between \SIrange{5}{30}{\watt} at these deep-UV wavelengths have been only demonstrated once before \cite{cooper2018}. They used \ce{SiO2} mirrors and observed rapid (\SI{<1}{\minute}) degradation of the mirrors at oxygen pressures below \SI{0.67}{\milli\bar}, attributing the result to hydrocarbon contamination versus surface oxygen depletion.  Using the \ce{CaF2} input coupler, our system cannot be affected by surface oxygen depletion. Therefore, it is reasonable to attribute the degradation of the \ce{CaF2} optics to hydrocarbon contamination. As seen in Fig. \ref{fig:O2a}, we observe degradation at \SI{1}{\milli\bar} for the same \ce{SiO2} input coupler used in \cite{cooper2018}.  

UV-ozone cleaning is a commonly used method to remove surface hydrocarbon contaminants \cite{KOHLI2019355}. With an \ce{O2} purge, the 244 nm intracavity radiation continuously generates ozone and atomic oxygen. Simultaneously, the 244 nm radiation also decomposes hydrocarbon contaminants into a variety of substances, including excited molecules and free radicals. The latter react with the atomic oxygen to form simpler volatile molecules, such as \ce{H2O}, \ce{CO2}, and \ce{NOx}, that are easily desorbed from the surface. The robustness of UV-ozone cleaning in our system is demonstrated clearly by the revival of the \ce{SiO2} input coupler in Fig. \ref{fig:UHVe}. 

Interestingly, as seen in Figs. \ref{fig:O2b}-\ref{fig:O2d} and \ref{fig:O2e}, with the \ce{CaF2} input coupler, we observe degradation-free enhancement on one hour time scales with up to four orders less oxygen maintained on the mirrors than was required for \ce{SiO2} in \cite{cooper2018}, and on four hour time scales with up to three orders of magnitude less oxygen. This suggests that the UV-ozone cleaning method is more effective for fluoride optics or there is an additional mechanism involved. UV conditioning of fluoride coatings is a well documented method to decrease their absorption \cite{eva1996laser, heber1998stability,heber1999changes}, and was clearly observed in our results (see Fig. \ref{fig:conditioning}). While the exact conditioning mechanism is not fully understood, Heber et. al. \cite{heber1998stability,heber1999changes} attribute it partially to radiation desorption of hydrocarbons on the surface, which they could not observe with oxide coatings. This desorption was observed in both air and argon atmospheres, suggesting hydrocarbon contaminants on fluoride coatings can be removed in oxygen-free or oxygen reduced environments---removing reliance on UV-ozone cleaning. This would explain why the \ce{CaF2} input coupler in our system can operate degradation free at significantly reduced \ce{O2} pressures. However, as observed in high vacuum, the radiation desorption cleaning method by itself is unable to fully prevent hydrocarbon contamination. This suggests the degradation free performance at low oxygen pressures relies on both cleaning methods. What remains unclear is the inability of UV-ozone cleaning to completely remove the effects hydrocarbon contamination in high vacuum over time, as seen in Fig. \ref{fig:UHVe}. Future work should try to characterize the surface contaminants present after this irreversible degradation, or whether the coating is damaged. 


\section{Summary}

In this work we have shown that relative to oxide-coated optics, a high-power, deep-UV CW enhancement cavity operating at \SI{244}{\nano\meter} in high vacuum with UV conditioned fluoride-coated optics, can have a degradation rate an order of magnitude smaller. This enables stable operation on hour time scales with up to \SI{10}{\watt} of intracavity power. To our knowledge, these are the highest CW deep-UV intracavity powers maintained on these time scales in high vacuum, and almost an order of magnitude higher than has been demonstrated with oxide-coated optics at these wavelengths. While combining oxygen with the deep-UV radiation can revive the optics after degradation, this method of revival is not sustainable.

However, with a continuous oxygen purge, we have shown that fluoride-coated optics can operate degradation-free with intracavity powers of up to \SI{20}{\watt} on multi-hour time scales with three to four orders of magnitude less oxygen than is required to prevent severe degradation with oxide-coated optics. Demonstration of such high powers on these time scales was enabled by the stability of our CW \SI{244}{\nano\meter} laser system, which can maintain Watt-level output powers on half-day time scales. 

Degradation of optics is a common hindrance to long term stability and power scaling of systems requiring deep-UV wavelengths. We have shown that fluoride optics are a promising alternative to oxide optics for development of a robust deep-UV CW enhancement cavity at increased intracavity powers and reduced reliance on UV-ozone cleaning. \\

\noindent\textbf{Funding}  \footnotesize{ERC consolidator grant (818053-Mu-MASS) and Swiss National Foundation grant 197346.}\\

\noindent\textbf{Acknowledgements} \footnotesize{The authors would like to thank Dylan Yost (CSU) for providing some of the optics used here as well as thoughtful discussions about the research. We also appreciate Samuel Cooper's (CSU) comments on this article as well as Michele Masseroni and Rebekka Garreis at ETH for their time helping us clean and diagnose the mirrors. We are extremely grateful to Konstantinos Simeonidis for his contributions to the laser system.} \\

\noindent\textbf{Disclosures} The authors declare no conflicts of interest. 








\bibliographystyle{unsrt}
\bibliography{sample}

\begin{thebibliography}{10}

\bibitem{savage2007ultraviolet}
Neil Savage.
\newblock Ultraviolet lasers.
\newblock {\em Nature Photonics}, 1(2):83--85, 2007.

\bibitem{lu2020deep}
Haoyuan Lu, Hao Xu, Jianye Zhao, and Dong Hou.
\newblock A deep ultraviolet mode-locked laser based on a neural network.
\newblock {\em Scientific reports}, 10(1):1--6, 2020.

\bibitem{smith2020low}
Callum~R Smith, Asbj{\o}rn Moltke, Abubakar~I Adamu, Mattia Michieletto,
  Patrick Bowen, Peter~M Moselund, Christos Markos, and Ole Bang.
\newblock Low-noise tunable deep-ultraviolet supercontinuum laser.
\newblock {\em Scientific Reports}, 10(1):1--11, 2020.

\bibitem{willenberg2020high}
Benjamin Willenberg, Fabian Brunner, Christopher~R Phillips, and Ursula Keller.
\newblock High-power picosecond deep-uv source via group velocity matched
  frequency conversion.
\newblock {\em Optica}, 7(5):485--491, 2020.

\bibitem{turcicova2019}
Hana Turcicova, Ondrej Novak, Lukas Roskot, Martin Smrz, Jiri Muzik, Michal
  Chyla, Akira Endo, and Tomas Mocek.
\newblock New observations on duv radiation at 257 nm and 206 nm produced by a
  picosecond diode pumped thin-disk laser.
\newblock {\em Optics express}, 27(17):24286--24299, 2019.

\bibitem{Burkley19}
Z.~Burkley, A.~D. Brandt, C.~Rasor, S.~F. Cooper, and D.~C. Yost.
\newblock Highly coherent, watt-level deep-uv radiation via a
  frequency-quadrupled yb-fiber laser system.
\newblock {\em Appl. Opt.}, 58(7):1657--1661, Mar 2019.

\bibitem{truppe2019spectroscopic}
Stefan Truppe, Silvio Marx, Sebastian Kray, Maximilian Doppelbauer, Simon
  Hofs{\"a}ss, Hanns~Christian Schewe, Nicole Walter, Jes{\'u}s
  P{\'e}rez-R{\'\i}os, Boris~G Sartakov, and Gerard Meijer.
\newblock Spectroscopic characterization of aluminum monofluoride with
  relevance to laser cooling and trapping.
\newblock {\em Physical Review A}, 100(5):052513, 2019.

\bibitem{zhao2017high}
Ruchen Zhao, Xiaohu Fu, Lei Zhang, Su~Fang, Jianfang Sun, Yan Feng, Zhen Xu,
  and Yuzhu Wang.
\newblock High-power continuous-wave narrow-linewidth 253.7 nm deep-ultraviolet
  laser.
\newblock {\em Applied optics}, 56(32):8973--8977, 2017.

\bibitem{cooper2018}
SF~Cooper, Z~Burkley, AD~Brandt, C~Rasor, and DC~Yost.
\newblock Cavity-enhanced deep ultraviolet laser for two-photon cooling of
  atomic hydrogen.
\newblock {\em Optics letters}, 43(6):1375--1378, 2018.

\bibitem{rasor2020laser}
C~Rasor and DC~Yost.
\newblock Laser-based measurement of parity violation in hydrogen.
\newblock {\em Physical Review A}, 102(3):032801, 2020.

\bibitem{zhang2020submicrosecond}
Chi Zhang, Fabian Pokorny, Weibin Li, Gerard Higgins, Andreas P{\"o}schl, Igor
  Lesanovsky, and Markus Hennrich.
\newblock Submicrosecond entangling gate between trapped ions via rydberg
  interaction.
\newblock {\em Nature}, 580(7803):345--349, 2020.

\bibitem{guardado2020quench}
Elmer Guardado-Sanchez, Benjamin Spar, Peter Schauss, Ron Belyansky, Jeremy~T
  Young, Przemyslaw Bienias, Alexey~V Gorshkov, Thomas Iadecola, and Waseem~S
  Bakr.
\newblock Quench dynamics of a fermi gas with strong long-range interactions.
\newblock {\em arXiv preprint arXiv:2010.05871}, 2020.

\bibitem{Oberthaler:2002qd}
M.~K. Oberthaler.
\newblock {Antimatter wave interferometry with positronium}.
\newblock {\em Nucl. Instrum. Meth. B}, 192:129--134, 2002.

\bibitem{parthey2011improved}
Christian~G Parthey, Arthur Matveev, Janis Alnis, Birgitta Bernhardt, Axel
  Beyer, Ronald Holzwarth, Aliaksei Maistrou, Randolf Pohl, Katharina Predehl,
  Thomas Udem, et~al.
\newblock Improved measurement of the hydrogen 1 s--2 s transition frequency.
\newblock {\em Physical review letters}, 107(20):203001, 2011.

\bibitem{ahmadi2018}
M~Ahmadi, BXR Alves, CJ~Baker, W~Bertsche, A~Capra, C~Carruth, CL~Cesar,
  M~Charlton, S~Cohen, R~Collister, et~al.
\newblock Characterization of the 1s--2s transition in antihydrogen.
\newblock {\em Nature}, 557(7703):71--75, 2018.

\bibitem{fleurbaey2018}
H{\'e}lene Fleurbaey, Sandrine Galtier, Simon Thomas, Marie Bonnaud, Lucile
  Julien, Fran{\c{c}}ois Biraben, Fran{\c{c}}ois Nez, Michel Abgrall, and
  Jocelyne Gu{\'e}na.
\newblock New measurement of the 1 s- 3 s transition frequency of hydrogen:
  contribution to the proton charge radius puzzle.
\newblock {\em Physical review letters}, 120(18):183001, 2018.

\bibitem{grinin2020}
Alexey Grinin, Arthur Matveev, Dylan~C Yost, Lothar Maisenbacher, Vitaly
  Wirthl, Randolf Pohl, Theodor~W H{\"a}nsch, and Thomas Udem.
\newblock Two-photon frequency comb spectroscopy of atomic hydrogen.
\newblock {\em Science}, 370(6520):1061--1066, 2020.

\bibitem{crivelli2018}
Paolo Crivelli.
\newblock The mu-mass (muonium laser spectroscopy) experiment.
\newblock {\em Hyperfine Interactions}, 239(1):1--9, 2018.

\bibitem{altiere2018}
Emily Altiere, Eric~R Miller, Tomohiro Hayamizu, David~J Jones, Kirk~W Madison,
  and Takamasa Momose.
\newblock High-resolution two-photon spectroscopy of a transition of xenon.
\newblock {\em Physical Review A}, 97(1):012507, 2018.

\bibitem{picker2017minuscule}
Ruediger Picker.
\newblock How the minuscule can contribute to the big picture: the neutron
  electric dipole moment project at triumf.
\newblock In {\em Proceedings of the 14th International Conference on
  Meson-Nucleon Physics and the Structure of the Nucleon (MENU2016)}, page
  010005, 2017.

\bibitem{kunz2000experimentation}
Roderick~R Kunz, Vladimir Liberman, and Deanna~K Downs.
\newblock Experimentation and modeling of organic photocontamination on
  lithographic optics.
\newblock {\em Journal of Vacuum Science \& Technology B: Microelectronics and
  Nanometer Structures Processing, Measurement, and Phenomena},
  18(3):1306--1313, 2000.

\bibitem{hollenshead2006}
Jeromy Hollenshead and Leonard Klebanoff.
\newblock Modeling radiation-induced carbon contamination of extreme
  ultraviolet optics.
\newblock {\em Journal of Vacuum Science \& Technology B: Microelectronics and
  Nanometer Structures Processing, Measurement, and Phenomena}, 24(1):64--82,
  2006.

\bibitem{gangloff2015}
Dorian Gangloff, Molu Shi, Tailin Wu, Alexei Bylinskii, Boris Braverman,
  Michael Gutierrez, Rosanna Nichols, Junru Li, Kai Aichholz, Marko Cetina,
  et~al.
\newblock Preventing and reversing vacuum-induced optical losses in
  high-finesse tantalum (v) oxide mirror coatings.
\newblock {\em Optics express}, 23(14):18014--18028, 2015.

\bibitem{cooper2020}
Samuel~F Cooper, Adam~D Brandt, Cory Rasor, Zakary Burkley, and Dylan~C Yost.
\newblock Cryogenic atomic hydrogen beam apparatus with velocity
  characterization.
\newblock {\em Review of Scientific Instruments}, 91(1):013201, 2020.

\bibitem{eva1996laser}
E~Eva, K~Mann, N~Kaiser, B~Anton, R~Henking, D~Ristau, P~Weissbrodt,
  D~Mademann, L~Raupach, and E~Hacker.
\newblock Laser conditioning of laf 3/mgf 2 dielectric coatings at 248 nm.
\newblock {\em Applied optics}, 35(28):5613--5619, 1996.

\bibitem{heber1999changes}
Joerg Heber, Roland Thielsch, Holger Blaschke, Norbert Kaiser, Uwe Leinhos, and
  A~G{\"o}rtler.
\newblock Changes in optical interference coatings exposed to 193-nm excimer
  laser radiation.
\newblock In {\em Laser-Induced Damage in Optical Materials: 1998}, volume
  3578, pages 83--96. International Society for Optics and Photonics, 1999.

\bibitem{burkley2017yb}
Z~Burkley, C~Rasor, SF~Cooper, AD~Brandt, and DC~Yost.
\newblock Yb fiber amplifier at 972.5 nm with frequency quadrupling to 243.1
  nm.
\newblock {\em Applied Physics B}, 123(1):1--6, 2017.

\bibitem{wu18}
Jingwei Wu, Xiushan Zhu, Hua Wei, Kort Wiersma, Michael Li, Jie Zong, Arturo
  Chavez-Pirson, Valery Temyanko, L.~J. LaComb, R.~A. Norwood, and
  N.~Peyghambarian.
\newblock Power scalable 10 w 976 nm single-frequency linearly polarized laser
  source.
\newblock {\em Opt. Lett.}, 43(4):951--954, Feb 2018.

\bibitem{drever1983laser}
RWP Drever, John~L Hall, FV~Kowalski, J\_ Hough, GM~Ford, AJ~Munley, and
  H~Ward.
\newblock Laser phase and frequency stabilization using an optical resonator.
\newblock {\em Applied Physics B}, 31(2):97--105, 1983.

\bibitem{briles2010simple}
Travis~C Briles, Dylan~C Yost, Arman Cing{\"o}z, Jun Ye, and Thomas~R Schibli.
\newblock Simple piezoelectric-actuated mirror with 180 khz servo bandwidth.
\newblock {\em Optics express}, 18(10):9739--9746, 2010.

\bibitem{chadi2013}
A.~Chadi, G.~Méjean, R.~Grilli, and D.~Romanini.
\newblock Note: Simple and compact piezoelectric mirror actuator with 100 khz
  bandwidth, using standard components.
\newblock {\em Review of Scientific Instruments}, 84(5):056112, 2013.

\bibitem{KOHLI2019355}
Rajiv Kohli.
\newblock Chapter 9 - applications of uv-ozone cleaning technique for removal
  of surface contaminants.
\newblock In Rajiv Kohli and K.L. Mittal, editors, {\em Developments in Surface
  Contamination and Cleaning: Applications of Cleaning Techniques}, pages
  355--390. Elsevier, 2019.

\bibitem{heber1998stability}
Joerg Heber, Roland Thielsch, Holger Blaschke, Norbert Kaiser, Klaus~R Mann,
  Eric Eva, Uwe Leinhos, and Andreas Goertler.
\newblock Stability of optical interference coatings exposed to low-fluence
  193-nm arf radiation.
\newblock In {\em Optical Microlithography XI}, volume 3334, pages 1041--1047.
  International Society for Optics and Photonics, 1998.

\end{thebibliography}
\end{document}